# Advancements in Signal Processing and Control Systems Using Z and L-Transforms

**A.Jeeva[1], Vijayabalan D[2], Maria Singaraj Rosary[2], Nasir Ali[3], and Fikadu Tesgera Tolasa[4,*]**

[1] Department of Mathematics, Panimalar Engineering College, Poonamalle, Chennai, India.
[2] Department of Mathematics, Veltech High-tech Dr. Rangarajan Dr. Sakunthala engineering college, Chennai, India.
[3] Department of Mathematics, COMSATS University Islamabad, Vehari Campus, Pakistan.
[4] Dambi Dollo University, Oromia, Ethiopia.

EMAIL ID'S: nasirzawar@gmail.com, drjeeva@veltech.edu.in , vijayabalantqb@gmail.com, roserosary19@gmail.com , mabdalla@kku.edu.sa , Nshat@kku.edu.sa , keebeekboonnii@gmail.com

Authors have received no specific funding or grant from any public or private sector.

**Abstract:** To excel in signal processing or control systems, a deep understanding of transforms is essential. But what exactly is this mathematical tool, and how does it function? In this article, we will explore the fundamentals of transforms, their properties, and their applications. Let's dive into one of the key concepts of modern signal processing. Additionally, we will discuss Z-transforms and integrated Z-transforms, highlighting their roles in signal processing. Moreover, to demonstrate the practical application of transforms, we will introduce a novel concept involving fuzzy random variables.

## 1. Introduction

A signal is any substance that transmits or transfers information. Signals include things like speaking, motion pictures, still images, music, and our heartbeat that we typically come into contact with in our daily lives. Anything that a signal transmits tends to be a function of an independent variable. The time, pressure, temperature, color intensity, spatial coordinates, and so on can all be regarded variables that are autonomous. The most commonly employed independent variable in signals is time, denoted by the letter "t." Amplitude is the value of a signal at any given value of the independent variable. A signal's waveform is an illustration or plot of the signal's amplitude as a function of a variable that is autonomous. Any signals can be mathematically characterized as a function of one or more independent variables. Any physical quantity that varies with one or more distinct variables is therefore defined as a signal. Any signal can be classified as continuous or discrete based on its number of sources, dimensions, and other characteristics. Analog signals are continually stated for all values of the independent variable. An analog signal is referred to as a continuous-time signal when time is its independent variable. Discrete signals are those that have discrete intervals of the independent variable defined in them. A discrete signal is called a discrete time signal when time is its independent variable. Before discussing the signal processing and control system; you must be an expert in it. Any device or procedure that receives an input signal and outputs it is considered a system. The relationship between the input and output signals is determined by the properties of the system. Signals are the inputs and outputs of a system. They could include any physical quantity that is subject to time-varying variations, such as voltage, current, pressure, temperature, or even information. Signals that the system processes carry information. The primary points of contact between a system and a signal are the input-output relationship, system features, signal processing, and system representation. Systems can be represented using a variety of models, including as transfer functions, state-space representations, and block diagrams. The structure, connections, and interactions between the input and output signals of the system are all made easier to understand with the aid of these representations. The connection between a system and a signal is intimately linked to the idea of transformation. The process of transforming an input signal into an output signal through a system's activity is known as transformation in the context of signal processing and systems. Signals can be altered using a variety of techniques, including frequency-domain, time-domain, invertible, non-linear, and frequency-domain transformations. It is critical to comprehend the signal transformations that can be applied to signals in system design and analysis in order to control and analyze signals to yield the intended results.

Various types of transforms, such as integral transforms, Laplace transforms, star transforms, Fourier transforms and discrete Fourier transforms, Zak transforms, and so on, are used in mathematics, physics, engineering, and other applied sciences, depending on whether the problem is discrete or continuous. Discrete systems cannot be investigated using the standard Laplace or Fourier transforms because they require continuous functions; alternatively, they can be easily described using the Z-transform. The $\mathscr{L}$-transform was first proposed by Laplace(1780) and then developed by Hurewicz as a controllable method of solving linear, constant-coefficient differential equations. In mathematical literature, the $\mathscr{L}$-transform concept is also known as the method of generating functions, which De-Moivre presented in probability theory. The $\mathcal{L}$-transform's most well-known uses are in signal processing, digital control, and geophysics, but it is also utilized extensively in all sectors of applied sciences and engineering. The Laplace transform used to transform a time domain signal to complex frequency domain in signals and systems,

The $\mathscr{L}$-transform of a continuous time signal $\mathcal{X}_t(n)$ is given by the equation,

$$\mathcal{L}\{\mathcal{X}_t(n)\} = \mathcal{X}_t(\mathcal{L}) = \int_{-\infty}^{\infty} e^{-sn}\, \mathcal{X}_t(n)\, dn$$

One of the most popular mathematical methods for converting data sequences from the domain to the complex frequency domain, usually in discrete time, is the $\mathcal{Z}$-transform. Originally conceived by W. Hurevicz(1947), the $\mathcal{Z}$-transform was subsequently designated as such in 1952 by the Colombia University sampled data control group, which included L.A. Zadeh, E.I. Jury, R.E. Kalman, and others, and was supervised by professor John R. Raggazini. It is essentially a discrete to Fourier transform (DFT) extension, but it has a few significant differences that make it very beneficial for digital signal processing. The independent variable $z$, which stands for the complex frequency domain, is the source of the name "$\mathcal{Z}$-transform." Simplifying the analysis and design of digital systems is its primary goal. One effective tool for digital systems is the $\mathcal{Z}$-transform. It is especially helpful for the analysis of liner time invariant (LTI) systems, which are extensively employed in audio processing, control systems, and telecommunication, among other fields. The $\mathcal{Z}$-transform provides a way to analyze the frequency response of a digital system, which is necessary for understanding its behavior and performance. From a strictly mathematical standpoint, the $\mathcal{Z}$-transform is essentially a power series representation of a discrete-time sequence, and so it exists when the series converges. However, because it is defined as the sum of an infinite number of items, it becomes very interesting to consider the Z-transform in relation to atypical numerical systems and non-standard theories that admit different types of countable (and uncountable) infinities and, in recent years, are experiencing a new impetus.

The $\mathcal{Z}$-transform of a discrete time signal $\mathcal{X}_t(n)$ is given by the equation,

$$\mathcal{Z}\{\mathcal{X}_t(n)\} = \mathcal{X}_t(\mathcal{Z}) = \sum_{n=-\infty}^{\infty} \mathcal{X}_t(n) z^{-n}$$

The numerator of the equation is the total of all the signal values scaled by $z$ raised to the power of their index, where $z$ is the complex frequency represented by the variable. We may use this method to obtain the complex-valued function $(z)$, which allows us to examine the signal's characteristics in the frequency domain.

The $\mathcal{Z}$-transform and $\mathscr{L}$-transform is helpful in digital signal processing because of its many significant features. The transform of a linear combination of signals is equal to the linear combination of each signal's unique transforms, making linearity one of the most significant characteristics. Because of this characteristic, complicated systems can be easily analyzed by disassembling them into more manageable parts. The Laplace transform, which is used to convert continuous-time data from the time domain to the complex frequency domain, is connected to the $\mathcal{Z}$ transform. A discrete-time signal's $\mathcal{Z}$ transform can be derived by evaluating the signal's Laplace transform at a particular point in s-space and replacing that result with $z$. A potent tool for analyzing continuous-time systems in many branches of research and engineering is the Laplace transform. It offers a method for examining a system's behaviuor in the frequency domain, which is crucial for comprehending its stability and performance. By utilizing the relationship between the $\mathcal{Z}$ and Laplace transforms, we may do unified analyses of both discrete- and continuous-time systems. Digital signal processing uses the $\mathcal{Z}$transform, a mathematical tool, to translate discrete-time signals into the frequency domain. With the help of this effective instrument, we may process signals by designing digital filters based on their frequency content analysis. The $\mathcal{Z}$-transform and $\mathscr{L}$-transform is a helpful tool because of its many qualities.

The basic idea of this article comes from the convolution theorem. Convolution theorem states that the transform of the convolution of two functions is equal to the product of the Fourier transforms of the individual functions. Mathematically, the convolution theorem can be stated as follows: Let $f(t)$ and $g(t)$ be two functions, and let $F(\omega)$ and $G(\omega)$ be their respective Fourier transforms. Then, the Fourier transform of the convolution of $f(t)$ and $g(t)$, denoted as $(f * g)(t)$, is equal to the product of $F(\omega)$ and $G(\omega) \Rightarrow (F * G)(\omega) = F(\omega) * G(\omega)$. This means that convolution in the time domain corresponds to multiplication in the frequency domain, and vice versa. The convolution theorem is widely used in various fields, such as signal processing, image processing, and linear systems theory, as it provides a powerful tool for simplifying and analyzing convolution operations. This allows us to apply the stochastic order application to the randomized signal with ease.

A few fundamental concepts of signals and systems are taken from references [2] and [7]. In contrast to traditional analysis, where the bilateral Z-transform usually does not exist anywhere, Caldarola F, Maiolo M, and Solferino V, [3] propose a completely novel type of $\mathcal{Z}$-transform of a complex sequence (or better yet, a family of infinitely many Z-transforms attached to the same sequence) whose existence is guaranteed almost everywhere on $C$. The method and technology employed to calculate the PTF of discrete digital filters are explained in Ciulla C[4]. The proof of concept has been effectively confirmed. This work introduces PTF-based two-dimensional $\mathcal{Z}$-space filtering. Adaptive fuzzy tracking control for a class of uncertain single-input/single-output nonlinear strict-feedback systems is the topic of Chen B, Liu X, Liu K and Lin C [5]. Unknown and desired control signals are directly simulated using fuzzy logic systems, and a unique direct adaptive fuzzy tracking controller is built via back stepping. The fundamental idea of the Laplace transforms its uses in signal processing and control systems, and the engineering applications of Laplace transforms were covered by Graf U [8]. A simplified and easy-to-understand approach of presenting a modified convolution and product theorem in the LCT domain derived via a quantum mechanical representation transformation is presented by Goel N, and Singh K [9]. When compared to the current convolution theorem for the *LCT*, it is discovered to be a more appropriate and superior proposal. Furthermore, utilizing the generated results, a filtering application is provided. After defining Huo H [10], we first present a novel idea of the canonical convolution operator and demonstrate its commutative, associative, and distributive characteristics, which might prove extremely beneficial in signal processing. Furthermore, it is demonstrated that the new canonical convolution operator linked to the *LCT* also satisfies the generalized Young's condition and the generalized convolution theorem. The proposed solution by Li Y, Li T, and Jing X [12] is studied through using fuzzy logic systems for dealing with unknown nonlinear functions and integrating adaptive fuzzy control design with adaptive backstepping technique. In particular, in the closed-loop system, the number of online learning parameters is decreased to 2n. A generalized convolution theorem for the *LCT* is proposed by Shi J, Liu X, and Zhang N[14], who also developed a similar product theorem related to the *LCT*. The ensuing results are shown to be special instances of the ordinary convolution theorem for the *FT*, the fractional convolution theorem for the *FRFT*, and several of the existing convolution theorems for the *LCT*. Additionally, a few uses for the deduced outcomes are showcased. The results of reduced adaptive fuzzy systems are found by Sun W, Su S, Wu Y., and Xia J [16]; these results are used to estimate the unknown function, which contains all of the system's state variables and guarantees that the backstepping design approach for nonstrict feedback nonlinear systems functions normally. A novel convolution structure for the *LCT* is introduced by Wei Ran, Q Li, and Ma [17]. It maintains the convolution theorem for the Fourier transform and is simple to apply when creating filters. We demonstrate that our obtained conclusions are special instances of some of the well-known theorems regarding the convolution theorem in the fractional Fourier transform *(FRFT)* domain and the *FT* domain. Based on an adaptive backstepping control method, Wu C, Liu J, Jing X, Li H, and Wu L. [18] address the problem of adaptive fuzzy control for a group of single-input, single-output nonlinear networked control systems with network-induced delay and data loss. Adaptive fuzzy tracking control for high-order nonlinear time-delay systems with full-state restrictions and input saturation is studied by Wu, Y., and Xie, X. [19]. Often utilized growth assumptions put on unknown system nonlinearities are eliminated by using a fuzzy approximation technique. Ze Ai, Huanqing Wang, and Haikuo Shen [20] study the issue of adaptive fuzzy fixed-time tracking control based on a class of nonlinear systems with unmodeled dynamics and dynamic disturbances.

## 2. Mathematical Properties and Applications of $\mathcal{Z}$ and $\mathscr{L}$-transforms

### 2.1 Linearity

The linearity property of the $\mathcal{Z}$-transform is one of the most important properties. It states that the $\mathcal{Z}$-transform is linear, which means that if we have two signals $\mathcal{X}_t(n)$ and $\mathcal{Y}_t(n)$ and their corresponding $\mathcal{Z}$-transforms $\mathcal{Z}\{\mathcal{X}_t(n)\}$ and $\mathcal{Z}\{\mathcal{Y}_t(n)\}$, then the $\mathcal{Z}$ transform of their sum $\mathcal{Y}_t\mathcal{X}_t(\mathcal{Z}) = \mathcal{Z}\{\mathcal{X}_t(n)\} + \mathcal{Z}\{\mathcal{Y}_t(n)\}$ is simply the sum of their individual $\mathcal{Z}$-transforms. This property is useful because it allows us to break down a complex signal into simpler signals that can be analyzed separately.

### 2.2 Time Shifting

The $\mathcal{Z}$-transform's time-shifting property states that a signal's $\mathcal{Z}$-transform is multiplied by $\mathcal{Z}^{-k}$if it is shifted by a specific amount, k. This applies to signals $\mathcal{X}_t(n)$. This characteristic is helpful because it enables us to examine how variations in time impact the signal's frequency domain. For instance, a signal's frequency content will shift to the left if we move it to the right in time.

**2.3 Scaling**

According to the $\mathcal{Z}$ transform's scaling property, a signal's $\mathcal{Z}$transform is multiplied by scalar α if it is multiplied by $\mathcal{X}_t\{1,2,3,\dots,n\}$, a scalar α. This feature is helpful since it lets us modify the signal's amplitude in the frequency domain. For instance, we may easily multiply a signal by a scalar in the time domain to increase a certain frequency component of the signal. This will multiply the corresponding frequency component in the frequency domain.

**2.4 Reversing Time**

According to the $\mathcal{Z}$-transform's time reversal property, the $\mathcal{Z}$-transform of $\mathcal{X}_t\{1,2,3,\dots,n\}$ is just mirrored along the complex plane if the samples' order is reversed. When evaluating symmetric signals and systems, this characteristic is helpful. For instance, we can make use of this attribute to streamline our analysis if our signal is symmetric about its middle.

**2.5 Convolution**

The $\mathcal{Z}$-transform of the convolution of two signals is just the product of their respective $\mathcal{Z}$-transforms, K($\mathcal{Z}$) and H($\mathcal{Z}$), according to the convolution property of the $\mathcal{Z}$-transform. When examining the frequency response of linear time-invariant systems, this characteristic is helpful. This feature can be used, for instance, to determine the frequency response of a system whose linear constant-coefficient-difference equation describes it. The above properties satisfied for $\boldsymbol{\mathcal{L}}$-transform also.

## 3. Applications of $Z$ Transform

The $\mathcal{Z}$-transform is a mathematical tool used to analyze and design discrete- time systems. It has a wide range of applications in various fields such as signal processing, control systems, and digital filters. Let's dive deeper into some of these applications.

**3.1 Signal Processing**

Analyzing and modifying signals with digital methods is called digital signal processing. An important tool in digital signal processing is the $\mathcal{Z}$-transform. Digital filters, which are employed for signal augmentation, noise reduction, and other purposes, are analyzed and designed using this technique. Digital filters are made to either boost desired frequencies or remove undesirable ones from a signal. We can evaluate a filter's frequency response and identify contemporary signal processing uses by using the $\mathcal{Z}$-transform.

**3.2 Control Systems**

Digital control system design and analysis also make use of the $\mathcal{Z}$-transform. Numerous industries, such as robotics, industrial automation, and aerospace, use digital control systems. With the $\mathcal{Z}$-transform, we may mimic a system's behavior in the frequency domain, examine a digital control system's stability, and create controllers with certain features. This creates and applies digital control mechanisms.

### 3.3 Digital Filters

$\mathcal{Z}$-transform are frequently used in digital filters. They are employed to boost desired frequencies or eliminate undesired ones from a transmission. The $\mathcal{Z}$-transform can be used to create filters with the required frequency response characteristics. The $\mathcal{Z}$-transform, for instance, can be used to examine the stability of the filter's low pass response. Because of this, it's a crucial tool for creating and using digital filters.

### 3.4 System Stability Analysis

In many applications, a system's stability is a crucial factor to take into account. If the output of a system stays bounded when the input is bounded, the system is stable. Digital control systems and digital filters are examples of discrete-time systems whose stability can be examined using the $\mathcal{Z}$-transform. The transfer function's poles in the z plane provide the basis for the first stability requirement. Designing and implementing stable systems requires the ability to examine a system's stability in the frequency domain, which the $\mathcal{Z}$-transform enables.

## 4. How z-transform working in signal processing and control systems

The $\mathcal{Z}$-transform plays a fundamental role in signal processing and control systems by providing a powerful tool for analyzing and manipulating discrete-time signals and systems. Here's how the $\mathcal{Z}$-transform works in these domains:

***Signal representation:*** We can express a discrete-time signal as a function of the complex variable $z$ using the $\mathcal{Z}$-transform. The $\mathcal{Z}$-transform $\mathcal{X}_t(\mathcal{Z})$ of a discrete-time signal $\mathcal{X}_t(n)$ is defined as the product of $\mathcal{X}_t(n)$ and $z^{-n}$ where n is the discrete time index. Analyzing the frequency content and signal behavior in the $z$-domain is possible with the $\mathcal{Z}$-transform .

***Frequency analysis****:* We can examine a discrete-time signal's frequency characteristics using the $\mathcal{Z}$-transform . We can ascertain the frequency response of the signal by assessing the $\mathcal{Z}$-transform at different locations on the complex plane. This aids in our comprehension of the signal's behavior at various frequencies and makes it possible for us to create systems and filters that alter or extract particular frequency components.

***System analysis and transfer functions:*** The $\mathcal{Z}$-transform can be used to analyze the frequency characteristics of a discrete-time signal. By evaluating the $\mathcal{Z}$-transform at various points on the complex plane, we may determine the signal's frequency response. This helps us understand how the signal behaves at different frequencies and enables us to design systems and filters that modify or extract specific frequency components.

***System design and frequency domain techniques:*** We can create digital filters and systems in the discrete-time domain thanks to the $\mathcal{Z}$-transform. We can create filters with desired frequency responses, such as low-pass, high-pass, or band-pass filters, by modifying the $\mathcal{Z}$-transform representation of a system. Additionally, the $\mathcal{Z}$-transform makes it easier to apply frequency domain methods to comprehend the stability and behavior of discrete-time systems, such as pole-zero analysis.

***Inverse Z-transform:*** A discrete-time signal can be recovered from its $\mathcal{Z}$-transform representation using the inverse $\mathcal{Z}$-transform, in the same way that a continuous-time signal can be recovered from its frequency representation using the inverse Fourier transform. We can recover the original discrete-time signal by going back from the $\mathcal{Z}$-domain to the time domain using the inverse $\mathcal{Z}$-transform.

***Transfer function manipulation:*** Transfer functions in the $\mathcal{Z}$-domain can be algebraically manipulated thanks to the $\mathcal{Z}$-transform. This enables the construction of controllers through the use of mathematical operations like addition, multiplication, and convolution, as well as the combining of various systems and system analysis. Complex control system design and analysis are made possible by these activities. To sum up, in signal processing and control systems, the $\mathcal{Z}$-transform offers a mathematical foundation for the analysis, design, and manipulation of discrete-time signals and systems. In addition to facilitating the application of numerous mathematical techniques in the $z$-domain, it permits frequency analysis, system representation, and system design.

## 5. Fresh Discoveries

To improve precision and lucidity, we will incorporate fuzzy logic into the Z-transforms. We introduce a novel use of Z-transforms for stochastic processes and fuzzy random variables.

### 5.1 Min and max function

The min and max functions are used in signal processing and control systems to find the lowest and highest values of a group of signals or variables. These functions are crucial for a number of tasks, including determining signal limits, identifying outliers, and locating peak values. The min and max functions in signal processing and control systems can be used in the following ways

### 5.2 Definition: Z-transform of fuzzy random variables

Let $x = (x_\alpha^L, x_\beta^U)$ be set of continuous non negative fuzzy random variables and $f(x)$ be a continuous function for all positive values of $Z$. A signal continuous in time taking continuous range of amplitude values defined for all times

$$Z\{(x_\alpha^L, x_\beta^U)(n)\} = X(z) = \sum_{n=-\infty}^{\infty} (x_\alpha^L, x_\beta^U)(n) z^{-n}$$

A discrete signal for which we only know values of the signal at discrete points in time

Provide the differential exists. It is denoted as $Z\{f\left(x_{\alpha}^{L}, x_{\beta}^{U}\right)(n)\} = -z\frac{d}{dx}\underline{F}\left(x_{\alpha}^{L}, x_{\beta}^{U}\right)$, where,

$X(z)$ Represents the Z-transform of the continuous and discrete time signal $\left(x_{\alpha}^{L}, x_{\beta}^{U}\right)(n)$.
$z^{-n}$ is the complex variable in the $Z$- domain domain.

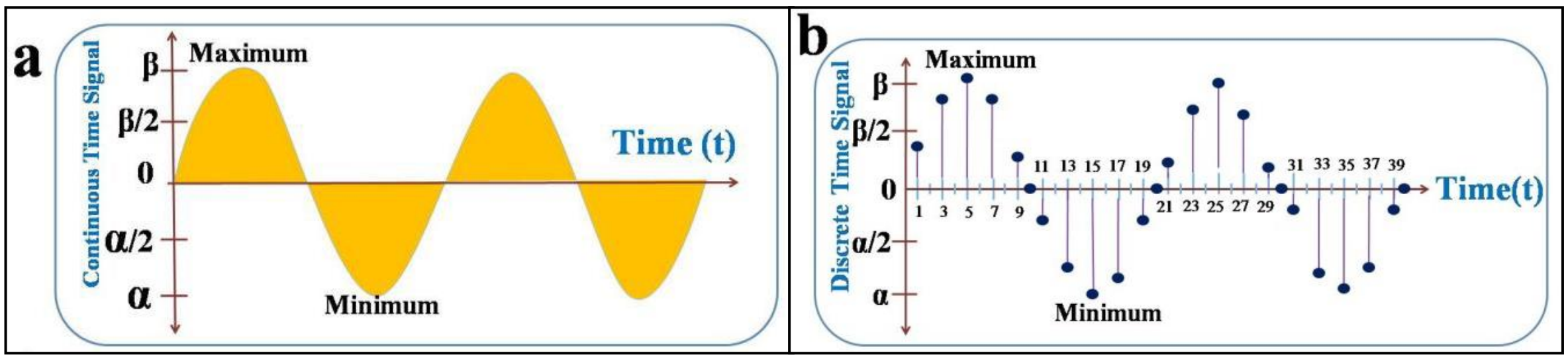

**Fig.1- (a) Continuous time signal; (b) Discrete Time signal.**

**5.3 Definition: $Z$-transform order of fuzzy random variables**

Let $X(z) and\, Y(z)$ be two absolutely continuous time non negative fuzzy random variables with probability density functions $f_i(x)\; and\; g_j(y)$and survival functions $\underline{F_i}(x)\; and\, \underline{F_j}(y)$. Hazard rate $H_i and H_j$ and mean residual time $M_i(t)\; and M_j(t)$. Then the fuzzy random variable of $z$-transforms,

$X$ is stochastically (Expectation) larger than $z$-transforms of $Y$ is denoted as $Z\{x(n)\} \leq_{ZTO}\; Z\{y(n)\}$and denoted by

$$\frac{\underset{\alpha\leq\beta\leq 1}{min}\sum_{n=-\infty}^{\infty} E(f_i\left(x_{\alpha}^{L}, x_{\beta}^{U}\right)(n)z^{-n})}{\underset{\alpha\leq\beta\leq 1}{min}\sum_{n=-\infty}^{\infty} E(g_j\left(y_{\alpha}^{L}, y_{\beta}^{U}\right)(n)z^{-n})} \geq 0 \qquad (1)$$

and

$$\frac{\underset{\alpha\leq\beta\leq 1}{max}\sum_{n=-\infty}^{\infty} E(f_i\left(x_{\alpha}^{L}, x_{\beta}^{U}\right)(n)z^{-n})}{\underset{\alpha\leq\beta\leq 1}{max}\sum_{n=-\infty}^{\infty} E(g_j\left(y_{\alpha}^{L}, y_{\beta}^{U}\right)(n)z^{-n})} \geq 0 \qquad (2)$$

$X$ is stochastically(Hazard rate) larger than $z$-transforms of $Y$ is denoted as
$Z\{x(n)\} \leq\geq_{I\&DHZTO} Z\{y(n)\}$and denoted by

$$\frac{\underset{\alpha\leq\beta\leq 1}{min}\sum_{n=-\infty}^{\infty}\{\frac{f_i\left(x_{\alpha}^{L}, x_{\beta}^{U}\right)}{\underline{F}_i\left(x_{\alpha}^{L}, x_{\beta}^{U}\right)}\}(n)z^{-n}}{\underset{\alpha\leq\beta\leq 1}{min}\sum_{n=-\infty}^{\infty}\{\frac{g_j\left(y_{\alpha}^{L}, y_{\beta}^{U}\right)}{\underline{G}_j\left(y_{\alpha}^{L}, y_{\beta}^{U}\right)}\}(n)z^{-n}} \geq 0 \qquad (3)$$

$$\frac{\underset{\alpha\leq\beta\leq 1}{max}\sum_{n=-\infty}^{\infty}\{\frac{f_i\left(x_{\alpha}^{L}, x_{\beta}^{U}\right)}{\underline{F}_i\left(x_{\alpha}^{L}, x_{\beta}^{U}\right)}\}(n)z^{-n}}{\underset{\alpha\leq\beta\leq 1}{max}\sum_{n=-\infty}^{\infty}\{\frac{g_j\left(y_{\alpha}^{L}, y_{\beta}^{U}\right)}{\underline{G}_j\left(y_{\alpha}^{L}, y_{\beta}^{U}\right)}\}(n)z^{-n}} \geq 0 \qquad (4)$$

$X$ is stochastically (Relative Hazard rate) larger than $z$-transforms of $Y$ is denoted as $Z\{x(n)\} \leq\geq_{I\&DRHZTO}\; Z\{y(n)\}$and denoted by

$$\frac{\min_{\alpha\le\beta\le 1}\sum_{n=-\infty}^{\infty}\frac{\frac{f_1(x_\alpha^L,x_\beta^U)}{\underline{F}_1(x_\alpha^L,x_\beta^U)}}{\frac{f_2(x_\alpha^L,x_\beta^U)}{\underline{F}_2(x_\alpha^L,x_\beta^U)}}(n)z^{-n}}{\min_{\alpha\le\beta\le 1}\sum_{n=-\infty}^{\infty}\frac{\frac{g_1(y_\alpha^L,y_\beta^U)}{\underline{G}_1(y_\alpha^L,y_\beta^U)}}{\frac{g_2(y_\alpha^L,y_\beta^U)}{\underline{G}_2(y_\alpha^L,y_\beta^U)}}(n)z^{-n}} \ge 0 \qquad (5)$$

and

$$\frac{\max_{\alpha\le\beta\le 1}\sum_{n=-\infty}^{\infty}\frac{\frac{f_1(x_\alpha^L,x_\beta^U)}{\underline{F}_1(x_\alpha^L,x_\beta^U)}}{\frac{f_2(x_\alpha^L,x_\beta^U)}{\underline{F}_2(x_\alpha^L,x_\beta^U)}}(n)z^{-n}}{\max_{\alpha\le\beta\le 1}\sum_{n=-\infty}^{\infty}\frac{\frac{g_1(y_\alpha^L,y_\beta^U)}{\underline{G}_1(y_\alpha^L,y_\beta^U)}}{\frac{g_2(y_\alpha^L,y_\beta^U)}{\underline{G}_2(y_\alpha^L,y_\beta^U)}}(n)z^{-n}} \ge 0 \qquad (6)$$

$X$ is stochastically(Likelihood ratio) larger than $z$-transforms of $Y$ is denoted as $Z\{x(n)\} \le \ge_{I\&DLZTO}$ $Z\{y(n)\}$and denoted by

$$\frac{\min_{\alpha\le\beta\le 1}\sum_{n=-\infty}^{\infty} f_i(x_\alpha^L,x_\beta^U)g_j(y_\alpha^L,y_\beta^U)(n)z^{-n}}{\min_{\alpha\le\beta\le 1}\sum_{n=-\infty}^{\infty} g_j(x_\alpha^L,x_\beta^U)f_i(y_\alpha^L,y_\beta^U)(n)z^{-n}} \ge 0 \qquad (7)$$

and

$$\frac{\max_{\alpha\le\beta\le 1}\sum_{n=-\infty}^{\infty} f_i(x_\alpha^L,x_\beta^U)g_j(y_\alpha^L,y_\beta^U)(n)z^{-n}}{\max_{\alpha\le\beta\le 1}\sum_{n=-\infty}^{\infty} g_j(x_\alpha^L,x_\beta^U)f_i(y_\alpha^L,y_\beta^U)(n)z^{-n}} \ge 0 \qquad (8)$$

- $X$ is stochastically (Mean residual life) larger than $z$-transforms of $Y$ is denoted as $Z\{x(n)\} \le_{I\&DMZTO}$ $Z\{y(n)\}$and denoted by

$$\frac{\min_{\alpha\le\beta\le 1}\sum_{n=1}^{\infty}(\frac{\int_0^\infty \left(\underline{F}_i(x_\alpha^L,x_\beta^U)(n)z^{-n}\right)dx}{\underline{F}_i(x_\alpha^L,x_\beta^U)}}{\min_{\alpha\le\beta\le 1}\sum_{n=1}^{\infty}(\frac{\int_0^\infty \left(\underline{G}_j(y_\alpha^L,y_\beta^U)(n)z^{-n}\right)dx}{\underline{G}_j(y_\alpha^L,y_\beta^U)})} \ge 0 \qquad (9)$$

and

$$\frac{\max_{\alpha\le\beta\le 1}\sum_{n=1}^{\infty}(\frac{\int_0^\infty \left(\underline{F}_i(x_\alpha^L,x_\beta^U)(n)z^{-n}\right)dx}{\underline{F}_i(x_\alpha^L,x_\beta^U)}}{\max_{\alpha\le\beta\le 1}\sum_{n=1}^{\infty}(\frac{\int_0^\infty \left(\underline{G}_j(y_\alpha^L,y_\beta^U)(n)z^{-n}\right)dx}{\underline{G}_j(y_\alpha^L,y_\beta^U)})} \ge 0 \qquad (10)$$

$X$ is stochastically(Relative Mean residual life) larger than $z$-transforms of $Y$ is denoted as $Z\{x(n)\} \le\ge_{I\&DRHZTO}$ $Z\{y(n)\}$and denoted by

$$\frac{\min_{\alpha\le\beta\le 1}\sum_{n=1}^{\infty}(\frac{\int_0^\infty \left(\underline{F_1}(x_\alpha^L,x_\beta^U)(n)z^{-n}\right)dx}{\underline{F_1}(x_\alpha^L,x_\beta^U)})}{\sum_{n=1}^{\infty}(\frac{\int_0^\infty \left(\underline{F_2}(x_\alpha^L,x_\beta^U)(n)z^{-n}\right)dx}{\underline{F_2}(x_\alpha^L,x_\beta^U)})} \le\ge_{I\&DMZTO}$$

and

$$\frac{\underset{\alpha\le\beta\le1}{max}\sum_{n=1}^{\infty}\left(\frac{\int_0^{\infty}\left(\underline{F_1}\left(x_\alpha^L,x_\beta^U\right)(n)z^{-n}\right)dx}{\underline{F_1}\left(x_\alpha^L,x_\beta^U\right)}\right)}{\sum_{n=1}^{\infty}\left(\frac{\int_0^{\infty}\left(\underline{F_2}\left(x_\alpha^L,x_\beta^U\right)(n)z^{-n}\right)dx}{\underline{F_2}\left(x_\alpha^L,x_\beta^U\right)}\right)} \le\ge_{I\&DMZTO}$$

$$\frac{\underset{\alpha\le\beta\le1}{max}\sum_{n=1}^{\infty}\left(\frac{\int_0^{\infty}\left(\underline{G}_1\left(y_\alpha^L,y_\beta^U\right)(n)z^{-n}\right)dx}{\underline{G}_2\left(y_\alpha^L,y_\beta^U\right)}\right)}{\sum_{n=1}^{\infty}\left(\frac{\int_0^{\infty}\left(\underline{G}_1\left(y_\alpha^L,y_\beta^U\right)(n)z^{-n}\right)dx}{\underline{G}_2\left(y_\alpha^L,y_\beta^U\right)}\right)} \qquad (12)$$

$X$ is stochastically(Aging intensity order) larger than $z$-transforms of $Y$ is denoted as $Z\{x(n)\} \le_{I\&DAIZTO} Z\{y(n)\}$and denoted by

$$\frac{\underset{\alpha\le\beta\le1}{min}\int_0^x\sum_{n=-\infty}^{\infty}f_i\left(x_\alpha^L,x_\beta^U\right)g_j\left(y_\alpha^L,y_\beta^U\right)(n)z^{-n}}{\underset{\alpha\le\beta\le1}{min}\int_0^y\sum_{n=-\infty}^{\infty}f_i\left(x_\alpha^L,x_\beta^U\right)g_j\left(y_\alpha^L,y_\beta^U\right)(n)z^{-n}} \ge 0 \qquad (13)$$

and

$$\frac{\underset{\alpha\le\beta\le1}{max}\int_0^x\sum_{n=-\infty}^{\infty}f_i\left(x_\alpha^L,x_\beta^U\right)g_j\left(y_\alpha^L,y_\beta^U\right)(n)z^{-n}}{\underset{\alpha\le\beta\le1}{max}\int_0^y\sum_{n=-\infty}^{\infty}f_i\left(x_\alpha^L,x_\beta^U\right)g_j\left(y_\alpha^L,y_\beta^U\right)(n)z^{-n}} \ge 0 \qquad (14)$$

**5.5 Theorem**

1. If $Z\{x(n)\} \le\ge_{I\&DRHZTO} Z\{y(n)\}$ are two non negative fuzzy random variable such that$Y$ is increasing or decreasing fuzzy Hazard rate order and $X$ is increasing or decreasing fuzzy Hazard rate order.
2. If $Z\{x(n)\} \le\ge_{I\&DRHZTO} Z\{y(n)\}$ are two non negative fuzzy random variable and $Z\{y(n)\} \le\ge_{I\&DHZTO} Z\{x(n)\}$, then $Z\{y(n)\} \le\ge_{I\&DLZTO} Z\{x(n)\}$.
3. If $Z\{x(n)\} \le\ge_{I\&DRHZTO} Z\{y(n)\}$ are two non negative fuzzy random variable implies that $Z\{x(n)\} \le_{I\&DAIZTO} Z\{y(n)\}$.
4. If $\underset{x,y\to0}{lim}\frac{g_j\left(y_\alpha^L,y_\beta^U\right)}{f_i\left(x_\alpha^L,x_\beta^U\right)} \le 1$,then $Z\{x(n)\} \le\ge_{I\&DRHZTO} Z\{y(n)\}$ implies

$Z\{x(n)\} \le\ge_{I\&DHZTO} Z\{y(n)\}$.

5. If $0 \le \underset{x,y\to0}{lim}\frac{g_j\left(y_\alpha^L,y_\beta^U\right)}{f_i\left(x_\alpha^L,x_\beta^U\right)} < \infty$, then $Z\{x(n)\} \le\ge_{I\&DRHZTO} Z\{y(n)\}$ implies that $Z\{x(n)\} \le\ge_{I\&DLZTO} Z\{y(n)\}$.

**Proof.** The proof of (1) is quite straightforward and hence omitted. The assertion (2) has been given in theorem in [11], to prove (3) first observe that, because of $Z\{x(n)\} \le\ge_{I\&DHZTO} Z\{y(n)\}$, We have

$$\underset{\alpha\le\beta\le1}{min}\sum_{n=-\infty}^{\infty}\frac{\frac{f_2\left(x_\alpha^L,x_\beta^U\right)}{\underline{F}_1\left(x_\alpha^L,x_\beta^U\right)}}{\frac{f_2\left(x_\alpha^L,x_\beta^U\right)}{\underline{F}_1\left(x_\alpha^L,x_\beta^U\right)}}(n)z^{-n}-\underset{\alpha\le\beta\le1}{min}\sum_{n=-\infty}^{\infty}\frac{\frac{g_2\left(y_\alpha^L,y_\beta^U\right)}{\underline{G}_1\left(y_\alpha^L,y_\beta^U\right)}}{\frac{g_2\left(y_\alpha^L,y_\beta^U\right)}{\underline{G}_1\left(y_\alpha^L,y_\beta^U\right)}}(n)z^{-n}\ge 0, \text{ for all } f\le g \quad (15)$$

Which is equivalent to

$$\min_{\alpha\le\beta\le1}\sum_{n=-\infty}^{\infty}\frac{\frac{f_2(x_\alpha^L,x_\beta^U)}{\underline{F}_2(x_\alpha^L,x_\beta^U)}}{\frac{f_2(y_\alpha^L,y_\beta^U)}{\underline{F}_2(y_\alpha^L,y_\beta^U)}}(n)z^{-n}-\min_{\alpha\le\beta\le1}\sum_{n=-\infty}^{\infty}\frac{\frac{g_1(y_\alpha^L,y_\beta^U)}{\underline{G}_1(y_\alpha^L,y_\beta^U)}}{\frac{g_1(x_\alpha^L,x_\beta^U)}{\underline{G}_1(x_\alpha^L,x_\beta^U)}}(n)z^{-n}\ge 0,\ \text{for all f} \le \text{g} \quad (16)$$

Therefore

$$\min_{\alpha\le\beta\le1}\int_0^{\infty}\sum_{n=-\infty}^{\infty}\frac{\frac{f_2(x_\alpha^L,x_\beta^U)}{\underline{F}_2(x_\alpha^L,x_\beta^U)}}{\frac{f_2(y_\alpha^L,y_\beta^U)}{\underline{F}_2(y_\alpha^L,y_\beta^U)}}(n)z^{-n}-\min_{\alpha\le\beta\le1}\sum_{n=-\infty}^{\infty}\frac{\frac{g_1(y_\alpha^L,y_\beta^U)}{\underline{G}_1(y_\alpha^L,y_\beta^U)}}{\frac{g_1(x_\alpha^L,x_\beta^U)}{\underline{G}_1(x_\alpha^L,x_\beta^U)}}(n)z^{-n}dx\ge 0, \quad (17)$$

This implies that $Z\{x(n)\}\le_{I\&DAIZTO} Z\{y(n)\}$. For proving (4), note that $\underline{G}_j(y_\alpha^L,y_\beta^U)(0)=\underline{F}_i(x_\alpha^L,x_\beta^U)(0)=1$. Since $Z\{x(n)\}\le\ge_{I\&DLZTO} Z\{y(n)\}$ implies*that*

$$\frac{\min_{\alpha\le\beta\le1}\sum_{n=-\infty}^{\infty}\{\frac{f_2(x_\alpha^L,x_\beta^U)}{\underline{F}_2(x_\alpha^L,x_\beta^U)}\}/\{\frac{f_2(y_\alpha^L,y_\beta^U)}{\underline{F}_2(y_\alpha^L,y_\beta^U)}\}(n)z^{-n}}{\min_{\alpha\le\beta\le1}\sum_{n=-\infty}^{\infty}\{\frac{g_1(y_\alpha^L,y_\beta^U)}{\underline{G}_1(y_\alpha^L,y_\beta^U)}\}/\{\frac{g_1(x_\alpha^L,x_\beta^U)}{\underline{G}_1(x_\alpha^L,x_\beta^U)}\}(n)z^{-n}}$$ is increasing in $(x,y)>0$, we have

$$\frac{\min_{\alpha\le\beta\le1}\sum_{n=-\infty}^{\infty}\{\frac{f_2(x_\alpha^L,x_\beta^U)}{\underline{F}_2(x_\alpha^L,x_\beta^U)}\}(n)z^{-n}}{\min_{\alpha\le\beta\le1}\sum_{n=-\infty}^{\infty}\{\frac{g_1(y_\alpha^L,y_\beta^U)}{\underline{G}_1(y_\alpha^L,y_\beta^U)}\}(n)z^{-n}}\ \lim_{x,y\to0}\left[\frac{\min_{\alpha\le\beta\le1}\sum_{n=-\infty}^{\infty}\{\frac{f_2(x_\alpha^L,x_\beta^U)}{f_2(y_\alpha^L,y_\beta^U)}\}/\{\frac{\underline{F}_2(x_\alpha^L,x_\beta^U)}{\underline{F}_2(y_\alpha^L,y_\beta^U)}\}(n)z^{-n}}{\min_{\alpha\le\beta\le1}\sum_{n=-\infty}^{\infty}\{\frac{g_1(y_\alpha^L,y_\beta^U)}{g_1(x_\alpha^L,x_\beta^U)}\}/\{\frac{\underline{G}_1(y_\alpha^L,y_\beta^U)}{\underline{G}_1(x_\alpha^L,x_\beta^U)}\}(n)z^{-n}}\right](n)z^{-n}$$

$$\le\lim_{x,y\to0}\sum_{n=1}^{\infty}\left\{\frac{g_j(x_\alpha^L,x_\beta^U)}{f_i(y_\alpha^L,y_\beta^U)}\right\}(n)z^{-n}\le 1, forallx>0, \quad (18)$$

Which is obviously gives $Z\{x(n)\}\le\ge_{I\&DHZTO} Z\{y(n)\}$. For the sake of proving (5), we observing that$\underline{G}_j(y_\alpha^L,y_\beta^U)(\infty)=\underline{F}_i(x_\alpha^L,x_\beta^U)(\infty)=0$, and using L'Hospital rule we get

$$\lim_{x,y\to0}\sum_{n=1}^{\infty}\left\{\frac{\underline{F}_i(x_\alpha^L,x_\beta^U)}{\underline{G}_i(y_\alpha^L,y_\beta^U)}\right\}(n)z^{-n}=\lim_{x,y\to0}\sum_{n=1}^{\infty}\left\{\frac{f_i(y_\alpha^L,y_\beta^U)}{g_j(x_\alpha^L,x_\beta^U)}\right\}(n)z^{-n}$$

$$=\lim_{x,y\to0}\sum_{n=1}^{\infty}(n)z^{-n}\left\{\frac{f_i(y_\alpha^L,y_\beta^U)}{g_j(x_\alpha^L,x_\beta^U)}\right\}^{-1}$$

Now since $Z\{x(n)\}\le\ge_{I\&DHZTO} Z\{y(n)\}$ we have

$$\frac{\min_{\alpha\le\beta\le1}\sum_{n=-\infty}^{\infty}\left\{\frac{g_j(x_\alpha^L,x_\beta^U)}{\underline{G}_j(x_\alpha^L,x_\beta^U)}\right\}(n)z^{-n}}{\min_{\alpha\le\beta\le1}\sum_{n=-\infty}^{\infty}\left\{\frac{f_i(y_\alpha^L,y_\beta^U)}{\underline{F}_i(y_\alpha^L,y_\beta^U)}\right\}(n)z^{-n}}\ge\lim_{x,y\to0}\frac{\min_{\alpha\le\beta\le1}\sum_{n=-\infty}^{\infty}\left\{\frac{g_j(x_\alpha^L,x_\beta^U)}{\underline{G}_j(x_\alpha^L,x_\beta^U)}\right\}(n)z^{-n}}{\min_{\alpha\le\beta\le1}\sum_{n=-\infty}^{\infty}\left\{\frac{f_i(y_\alpha^L,y_\beta^U)}{\underline{F}_i(y_\alpha^L,y_\beta^U)}\right\}(n)z^{-n}} \quad (19)$$

$$=\lim_{x,y\to0}\min_{\alpha\le\beta\le1}\sum_{n=-\infty}^{\infty}\left\{\frac{g_j(y_\alpha^L,y_\beta^U)}{f_i(x_\alpha^L,x_\beta^U)}\right\}(n)z^{-n}\min_{\alpha\le\beta\le1}\sum_{n=-\infty}^{\infty}\left\{\frac{\underline{F}_i(x_\alpha^L,x_\beta^U)}{\underline{G}_j(y_\alpha^L,y_\beta^U)}\right\}(n)z^{-n}$$

$$=1.$$

That is $Z\{y(n)\} \leq\geq_{I\&DHZTO}\ Z\{x(n)\}$, this in view of assertion (2) implies that $Z\{y(n)\} \leq\geq_{I\&DLZTO}\ Z\{x(n)\}$.
The above results holds true for maximum function also.

**5.6 Theorem**

1. If $Z\{x(n)\} \leq\geq_{I\&DRMRZTO}\ Z\{y(n)\}$ are two non negative fuzzy random variable such that $Y$ is increasing or decreasing fuzzy Relative mean residual life order and $X$ is increasing or decreasing fuzzy Relative mean residual life order
2. If $Z\{x(n)\} \leq\geq_{I\&DRMRZTO} Z\{y(n)\}$ are two non negative fuzzy random variable and $Z\{y(n)\} \leq\geq_{I\&DMRZTO} Z\{x(n)\}$ then $Z\{y(n)\} \leq\geq_{I\&DHRZTO} Z\{x(n)\}$ .
3. If $0 < \lim_{x,y\to 0} \frac{g_j\left(y_\alpha^L, y_\beta^U\right)}{f_i\left(x_\alpha^L, x_\beta^U\right)} \leq \infty$,then $Z\{x(n)\} \leq\geq_{I\&DRMZTO}\ Z\{y(n)\}$ implies that $Z\{y(n)\} \leq\geq_{I\&DHZTO}\ Z\{x(n)\}$.
4. If $E(Z\{x(n)\} \leq\geq E(Z\{y(n)\})$ are two non negative fuzzy random variable of $Z$transform then $Z\{x(n)\} \leq\geq_{I\&DRMRZTO} Z\{y(n)\}$the $Z\{y(n)\} \leq \geq_{I\&DHRZTO} Z\{x(n)\}$implies that $Z\{x(n)\} \leq\geq_{I\&DMRZTO} Z\{y(n)\}$.

**Proof.** The proof of (1) and (2) are easily obtained and hence omitted. To prove (3) by using L' Hospital rule we have

$$\lim_{x,y\to 0} \frac{\min_{\alpha\leq\beta\leq 1} \sum_{n=1}^{\infty} \left(\frac{\int_0^\infty \left(\underline{F}_1\left(x_\alpha^L, x_\beta^U\right)(n) z^{-n}\right)dx}{\underline{F}_1\left(x_\alpha^L, x_\beta^U\right)}\right)}{\min_{\alpha\leq\beta\leq 1} \sum_{n=1}^{\infty} \left(\frac{\int_0^\infty \left(\underline{G}_2\left(y_\alpha^L, y_\beta^U\right)(n) z^{-n}\right)}{\underline{G}_2\left(y_\alpha^L, y_\beta^U\right)}\right)} = \lim_{x,y\to 0} \frac{\min_{\alpha\leq\beta\leq 1} \sum_{n=1}^{\infty} \frac{\left(\underline{F}_2\left(x_\alpha^L, x_\beta^U\right)(n) z^{-n}\right)}{\underline{F}_2\left(x_\alpha^L, x_\beta^U\right)}}{\min_{\alpha\leq\beta\leq 1} \sum_{n=1}^{\infty} \frac{\left(\underline{G}_1\left(y_\alpha^L, y_\beta^U\right)(n) z^{-n}\right)}{\underline{G}_1\left(y_\alpha^L, y_\beta^U\right)}}$$

$$= \lim_{x,y\to 0} \frac{\min_{\alpha\leq\beta\leq 1} \sum_{n=1}^{\infty} \frac{\left(\underline{G}_1\left(y_\alpha^L, y_\beta^U\right)(n) z^{-n}\right)}{\underline{G}_1\left(y_\alpha^L, y_\beta^U\right)}}{\min_{\alpha\leq\beta\leq 1} \sum_{n=1}^{\infty} \frac{\left(\underline{F}_2\left(x_\alpha^L, x_\beta^U\right)(n) z^{-n}\right)}{\underline{F}_2\left(x_\alpha^L, x_\beta^U\right)}}$$

Now, let us observe that

$$\min_{\alpha\leq\beta\leq 1} \sum_{n=1}^{\infty} \frac{\frac{\int_0^\infty \left(\underline{F}_i\left(x_\alpha^L, x_\beta^U\right)(n) z^{-n}\right)dx}{\underline{F}_i\left(x_\alpha^L, x_\beta^U\right)}}{\frac{\int_0^\infty \left(\underline{G}_j\left(y_\alpha^L, y_\beta^U\right)(n) z^{-n}\right)dx}{\underline{G}_j\left(y_\alpha^L, y_\beta^U\right)}} \leq \frac{\min_{\alpha\leq\beta\leq 1} \sum_{n=1}^{\infty} \left(\frac{\int_0^\infty \left(\underline{F}_i\left(x_\alpha^L, x_\beta^U\right)(n) z^{-n}\right)dx}{\underline{F}_i\left(x_\alpha^L, x_\beta^U\right)}\right)}{\min_{\alpha\leq\beta\leq 1} \sum_{n=1}^{\infty} \left(\frac{\int_0^\infty \left(\underline{G}_j\left(y_\alpha^L, y_\beta^U\right)(n) z^{-n}\right)dx}{\underline{G}_j\left(y_\alpha^L, y_\beta^U\right)}\right)}$$

$$= \lim_{x,y\to 0} \frac{\min_{\alpha\leq\beta\leq 1} \sum_{n=1}^{\infty} \left(\frac{\int_0^\infty \left(\underline{F}_1\left(x_\alpha^L, x_\beta^U\right)(n) z^{-n}\right)dx}{\underline{F}_1\left(x_\alpha^L, x_\beta^U\right)}\right)}{\min_{\alpha\leq\beta\leq 1} \sum_{n=1}^{\infty} \left(\frac{\int_0^\infty \left(\underline{G}_2\left(y_\alpha^L, y_\beta^U\right)(n) z^{-n}\right)}{\underline{G}_2\left(y_\alpha^L, y_\beta^U\right)}\right)} \lim_{x,y\to 0} \frac{\min_{\alpha\leq\beta\leq 1} \sum_{n=1}^{\infty} \frac{\left(\underline{G}_1\left(y_\alpha^L, y_\beta^U\right)(n) z^{-n}\right)}{\underline{G}_1\left(y_\alpha^L, y_\beta^U\right)}}{\min_{\alpha\leq\beta\leq 1} \sum_{n=1}^{\infty} \frac{\left(\underline{F}_2\left(x_\alpha^L, x_\beta^U\right)(n) z^{-n}\right)}{\underline{F}_2\left(x_\alpha^L, x_\beta^U\right)}}$$

$$= \frac{\underset{x,y\to 0}{lim} \frac{\underset{\alpha\le\beta\le1}{min} \sum_{n=1}^{\infty} \frac{\left(\underline{G}_1\left(y_\alpha^L, y_\beta^U\right)(n)z^{-n}\right)}{\underline{G}_1\left(y_\alpha^L, y_\beta^U\right)}}{\underset{\alpha\le\beta\le1}{min} \sum_{n=1}^{\infty} \frac{\left(\underline{F}_2\left(x_\alpha^L, x_\beta^U\right)(n)z^{-n}\right)}{\underline{F}_2\left(x_\alpha^L, x_\beta^U\right)}}}{}$$

$$* \underset{x,y\to 0}{lim} \frac{\underset{\alpha\le\beta\le1}{min} \sum_{n=1}^{\infty} \frac{\left(\underline{G}_1\left(y_\alpha^L, y_\beta^U\right)(n)z^{-n}\right)^{-1}}{\underline{G}_1\left(y_\alpha^L, y_\beta^U\right)^{-1}}}{\underset{\alpha\le\beta\le1}{min} \sum_{n=1}^{\infty} \frac{\left(\underline{F}_2\left(x_\alpha^L, x_\beta^U\right)(n)z^{-n}\right)^{-1}}{\underline{F}_2\left(x_\alpha^L, x_\beta^U\right)^{-1}}} = 1$$

for all x, y ≥ 0

Therefore we conclude that $Z\{x(n)\} \le\ge_{I\&DMRZTO} Z\{y(n)\}$, which by results of assertion (2) implies that $Z\{y(n)\} \le\ge_{I\&DHRZTO} Z\{x(n)\}$.

To prove (4), for all $x, y \ge 0$ we have

$$\underset{\alpha\le\beta\le1}{min} \sum_{n=1}^{\infty} \frac{\frac{\int_0^{\infty} \left(\underline{F}_i\left(x_\alpha^L, x_\beta^U\right)(n)z^{-n}\right) dx}{\underline{F}_i\left(x_\alpha^L, x_\beta^U\right)}}{\frac{\int_0^{\infty} \left(\underline{G}_j\left(y_\alpha^L, y_\beta^U\right)(n)z^{-n}\right) dx}{\underline{G}_j\left(y_\alpha^L, y_\beta^U\right)}} \ge \underset{\alpha\le\beta\le1}{min} \sum_{n=1}^{\infty} \frac{\frac{\int_0^{\infty} \left(\underline{F}_i(0)(n)z^{-n}\right) dx}{\underline{F}_i(0)}}{\frac{\int_0^{\infty} \left(\underline{G}_j(0)(n)z^{-n}\right) dx}{\underline{G}_j(0)}}$$

$$= \underset{\alpha\le\beta\le1}{min} \sum_{n=-\infty}^{\infty} \frac{E\left(f_i\left(x_\alpha^L, x_\beta^U\right)(n)z^{-n}\right)}{E\left(g_j\left(y_\alpha^L, y_\beta^U\right)(n)z^{-n}\right)} \ge 1$$

The above results hold true for maximum function also.

**5.7 *Z*-transform algorithm of continuous signal** $\left(x_\alpha^L, x_\beta^U\right)(n)$

We know that

$$Z\left\{\left(x_\alpha^L, x_\beta^U\right)\right\} = \sum_{n=\infty}^{\infty} \left(x_\alpha^L, x_\beta^U\right)(n)z^{-n}$$

$$= \sum_{n=1}^{\infty} \left(x_\alpha^L, x_\beta^U\right)^n$$

$$= \sum_{n=1}^{\infty} \left(x_\alpha^L, x_\beta^U\right)^n \frac{1}{z^n}$$

$$= \sum_{n=1}^{\infty} \frac{\left(x_\alpha^L, x_\beta^U\right)^n}{z^n}$$

$$= \sum_{n=1}^{\infty} \left(\frac{\left(x_\alpha^L, x_\beta^U\right)}{z}\right)^n$$

$$= 1 + \frac{\left(x_\alpha^L, x_\beta^U\right)}{z} + \left(\frac{\left(x_\alpha^L, x_\beta^U\right)}{z}\right)^2 + \left(\frac{\left(x_\alpha^L, x_\beta^U\right)}{z}\right)^3 + \cdots$$

$$= \left(\frac{z - \left(x_\alpha^L, x_\beta^U\right)}{z}\right)^{-1}$$

$$Z\{\left(x_\alpha^L, x_\beta^U\right)\} = \left(\frac{z}{z-\left(x_\alpha^L, x_\beta^U\right)}\right)^1, |z| > /\left(x_\alpha^L, x_\beta^U\right)/ \quad (20)$$

**5.8 Example**

Find the $Z$-transform of the sequence signal of fuzzy random variable,

$$\left(x_\alpha^L, x_\beta^U\right)(n) = \{0.2,0.4,0.6,0.8,0.1\}$$

**Solution**

Given that $\left(x_\alpha^L, x_\beta^U\right)(n) = \{0.2,0.4,0.6,0.8,0.1\}$.

Substituting the sequence values in the above equation, we obtain,

$\left(x_\alpha^L, x_\beta^U\right)(n)z^{-n} = 0.2z^5 + 0.4z^4 + 0.6z^3 + 0.8z^2 + 1z^1$ (by eqn.20)

The $\left(x_\alpha^L, x_\beta^U\right)(n)z^{-n}$.

converges for all values of $z$ except at $z = 0$ *and* $z = 1$.

## 6. $Z$-transform's integrability

Let $\{\mathcal{X}_t(n_1), \mathcal{X}_t(n_2), \mathcal{X}_t(n_2), \dots, \mathcal{X}_t(n_m)\}$ be a sequence of $\mathcal{Z}$-transform defined for $\{m = 1,2,3, \dots, \infty\}$ and$\mathcal{X}_t(n_m) = 0$, then its $Z$-transform is defined to be

$$\int_{m=1}^{\infty} \mathcal{Z}\{\mathcal{X}_t(n_\alpha^L, n_\beta^U)\}dx = \int_{m=1}^{\infty} \sum_{n=1}^{\infty} \mathcal{X}_t(n_\alpha^L, n_\beta^U)z^{-n}dx$$

Where, $z$ is arbitrary constant. The integrability of $\mathcal{Z}$-transform is a mathematical powerful tool used in core dimensional signal processing to relate a discrete-time and continuous signal in two or more dimensions to the complex frequency domain. It is a generalization of the one-dimensional $Z$-transform and is useful for analyzing the different signal and frequency content of signals in multiple dimensions.

### 6.1 Stochastic order of $\mathcal{Z}$-transform's integrability

To compare signals and control systems using the $\mathcal{Z}$-transform, you need to understand how the $\mathcal{Z}$-transform can represent signals and systems in the discrete-time domain. Here's a step-by-step guide on how to perform the comparison. Apply the $\mathcal{Z}$-transform to the given signal to obtain its representation in the $\mathcal{Z}$-domain. This step involves expressing the signal as a sequence$\mathcal{X}_t(n_\alpha^L, n_\beta^U)$ and evaluating the $\mathcal{Z}$-transform using the formula mentioned above.

**Obtain the $\mathcal{Z}$-transform of the control system:**

If you have a discrete-time control system, you need to obtain its transfer function and then apply the $\mathcal{Z}$-transform to it. The transfer function typically represents the relationship between the input and output of a control system. Once you have the transfer function, substitute $z$ with the complex variable $z^{-1}$to obtain the $\mathcal{Z}$-transform of the control system.

**Compare the $\mathcal{Z}$-transforms:**

After obtaining the $\mathcal{Z}$-transforms of both the signal and the control system, you can compare them to analyze their properties. This comparison can involve various techniques, such as determining the poles and zeros, evaluating stability, analyzing frequency response, and examining other characteristics.

**Pole-zero analysis**: Analyze the location of poles and zeros in the $\mathcal{Z}$-plane to determine stability and system response. The following Figure 2, shown pole zero analysis of Z-transform under the fuzzy environment.

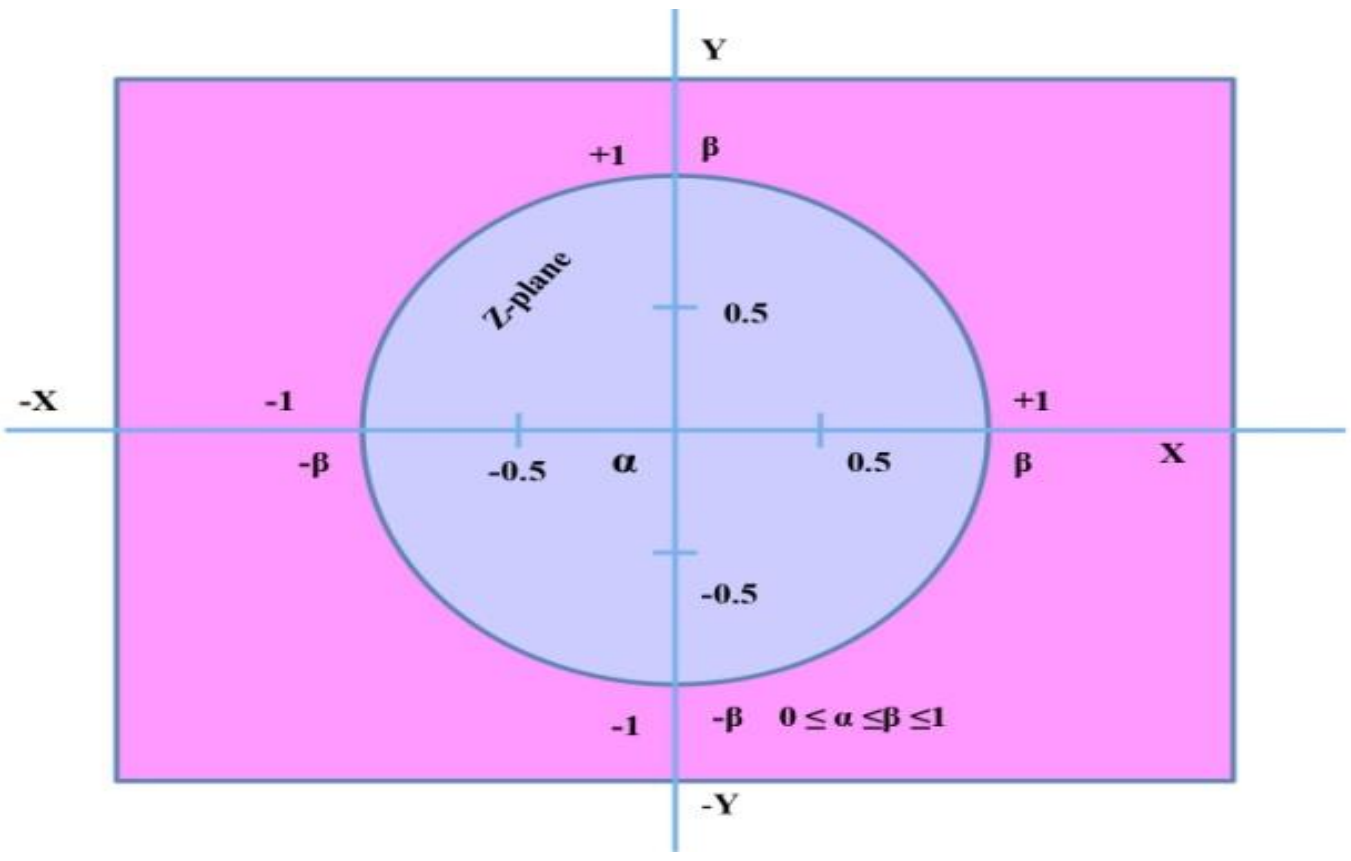

*Figure 2: pole zero analysis of $\mathcal{Z}$-transform*

**Frequency response:** Evaluate the $\mathcal{Z}$transform at specific frequencies to analyze the system's behavior in the frequency domain.

**System identification:** Compare the $\mathcal{Z}$-transforms to identify the system parameters, such as damping ratio, natural frequency, etc.

Remember that comparing signals and control systems using the $\mathcal{Z}$-transform is a complex process, and it requires a good understanding of the underlying theory and mathematics. It's recommended to consult textbooks, academic resources, or seek expert guidance to ensure accurate analysis and interpretation.

Let $\mathcal{Z}\{\mathcal{X}_t(n_\alpha^L, n_\beta^U)\}$ $=\{\mathcal{X}_t(n_1), \mathcal{X}_t(n_2), \mathcal{X}_t(n_2), \dots, \mathcal{X}_t(n_m)\}$ and

$\mathcal{Z}\{\mathcal{Y}_t(n_\alpha^L, n_\beta^U)\}$ =$\{\mathcal{Y}_t(n_1), \mathcal{Y}_t(n_2), \mathcal{Y}_t(n_2), \dots, \mathcal{Y}_t(n_m)\}$ be a sequence of fuzzy random variable, defined for $\{m = 1,2,3, \dots, \infty\}$ and $x_n(n) = 0, y_n(n) = 0$ then its $\mathcal{Z}$-transform $X$ is stochastically dominant than $Y$ is denoted by, $\mathcal{Z}\{\mathcal{X}_t(n_\alpha^L, n_\beta^U)\} \leq_{SD} \mathcal{Z}\{\mathcal{Y}_t(n_\alpha^L, n_\beta^U)\}$

$$\int_1^n \sum_{-\infty}^{\infty} \{\mathcal{X}_t(n_\alpha^L, n_\beta^U)\} z^{-n} dx \leq_{SD} \int_1^n \sum_{-\infty}^{\infty} \{\mathcal{X}_t(n_\alpha^L, n_\beta^U)\} z^{-n} dx$$

## 7. Comparative Study

Recently, $\mathcal{Z}$-transforms have been playing an important role in understanding signal processing and control systems. More researchers came up with original solutions to the fundamental issues in control systems and signal processing. The creation of fractional Z-transform algorithms and their uses in control, system identification, and signal processing have been the main topics of recent study. $\mathcal{Z}$-transform-based time-frequency analysis techniques have been investigated recently for speech processing, picture processing, and biological signal analysis, among other applications. Studies have been carried out to create sparse Z-transform algorithms that take advantage of signal sparsity to accomplish effective signal processing and representation in the $z$ domain. The goal of multidimensional $\mathcal{Z}$-transform research is to create effective algorithms for multidimensional $\mathcal{Z}$-transform computation and investigate its uses in compression, analysis, and image and video processing. The goal of applications in data science and machine learning research is to investigate the use of $\mathcal{Z}$-transform -based techniques for forecasting, anomaly detection, time series classification, and feature extraction. Now this research is described absolutely new innovation idea of signal processing and control system. This article explained the signal's lifetime. Determining the signal lifetime is a critical task in communication engineering. We can quickly determine a modest level signal loss in an electrical circuit using this method. Prospective researchers can quickly and easily determine where there is signal loss in the circuit and make the necessary adjustments.

## Conclusion

An effective mathematical tool for digital signal processing and control systems analysis is the $Z$ transform. It makes the study and design of digital systems easier by enabling the transformation of signals from the time domain to the frequency domain. Some of the most important characteristics of the $Z$transforms that we have examined are convolution, linearity, scaling, time shifting, and time reversal. A few of its uses, such as signal processing, control systems, digital filters, and system stability analysis, have also been investigated. One of your most crucial tools going forward will surely be the Z transform as you continue to study signal processing. Furthermore, $Z$ is an effective instrument with a broad range of applications in numerous fields. It enables us to study and create control systems, stability systems, and digital filters. Its significance in contemporary technology cannot be emphasized, and it will be a vital tool for many years to come.

**Declaration:**

- Availability of data and materials: The data is provided on request to the authors.
- Conflicts of interest: The authors declare that they have no conflicts of interest and all the agree to publish this paper under academic ethics.

**References:**

[1] Abouammoh. A.M, Abdulghani. S. A and Qamber. I.S, “On partial orderings and testing of new better than renewal used classes,” Reliability *Engineering and System Saftey,* vol. 43, no.1 pp.37-41,(1994).

[2] Bilinear transform. https://en.wikipedia.org/wiki/Bilinear_transform. Accessed 21 Aug (2021).

[3] Caldarola. F, Maiolo. M, Solferino. V: “A new approach to the Z-transform through infinite computation”. Commun. Nonlinear Sci. Numer. Simul. **82**, 105019 (2020).

[4] Ciulla. C :Two-dimensional Z-space filtering using pulse-transfer function. Circuits Syst. Signal Process. **42**, 255-276 (2023).

[5] Chen. B, Liu. X, Liu. K, Lin. C, Direct adaptive fuzzy control of nonlinear strict-feedback systems. *Automatica*. 2009; **45**(6): 1530-1535.

[6] Di Crescenzo. A, “Dual Stochastic Orderings describing ageing Properties of devices of unknown age,” Communications in Statistics-Stochastics Models, vol.15,no. 3,pp, 561-576,(1999).

[7] Engelberg. S “Digital signal processing. An experimental approach”.(2008).

[8] Graf. U “Applied Laplace transforms and z-transforms for scientists and engineers’’: a Computational approach using a mathematica package.(2004).

[9] Goel. N, Singh. K: “A modified convolution and product theorem for the linear canonical transform derived by representation transformation in quantum mechanics”. Int. J. Appl. Math. Comput. Sci. **23**, 685–695 (2013).

[10] Huo. H: A new convolution theorem associated with the linear canonical transform. SIViP **13**, 127–133 (2019).

[11] Lai. C.D and Xie. M, Stochastic Ageing and Dependence for Reliability, Springer, New York, NY,USA,(2006).

[12] Li. Y, Li. T, Jing. X, Indirect adaptive fuzzy control for input and output constrained nonlinear systems using a barrier Lyapunov function. *Int J Adapt Control Signal Process*. ; **28**(2): 184-199.(2014)

[13] Nanda. A.K and Kundu. A, "On generalized stochastic orders of dispersion- type," Calcutta *Statistical Association Bulletin*, Vol. 61, pp. 155-182, (2009).

[14] Shi. J, Liu. X, Zhang. N: Generalized convolution and product theorems associated with linear canonicl transform. SIViP **8**, 967–974 (2014)

[15] Shaked. M and Shanthikumar. J. G, Stochastic Orders, Springers, NEW York, NY.USA,(2007).

[16] Sun. W, Su. S, Wu. Y, Xia. J, Novel adaptive fuzzy control for output constrained stochastic nonstrict feedback nonlinear systems. *IEEE Trans Fuzzy Syst*. **29**(5): 1188-1197.(2021)

[17] Wei. D, Ran. Q, Li. Y, Ma. J, Tan .L, "A convolution and product theorem for the linear canonical transform", . Lett. **16**, 853–856 (2009).

[18] Wu. C, Liu. J, Jing. X, Li. H, Wu. L, "Adaptive fuzzy control for nonlinear networked control systems". *IEEE Trans Syst Man Cybern Syst*. ; **47**(8): 2420-2430.(2017)

[19] Wu. Y, Xie. X, "Adaptive fuzzy control for high-order nonlinear time-delay systems with full-state constraints and input saturation". *IEEE Trans Fuzzy Syst* ; **28**(8): 1652-1663.(2020)

[20] Ze Ai, Huanqing Wang, Haikuo Shen "Adaptive fuzzy fixed-time tracking control of nonlinear systems with unmodeled dynamics", *international journal adoptive control and signal processing*, Volume 38, Issue 6 , 2024.